%% file: main.tex
\documentclass[sigconf]{acmart}
\usepackage[utf8]{inputenc}
\usepackage{graphicx}
\usepackage{subcaption}
\usepackage{caption}
\usepackage{listings}
\usepackage{float}
\usepackage{enumitem}

\lstdefinestyle{XMLStyle}{
    language=XML,
    basicstyle=\small\ttfamily,
    frame=single,
    breaklines=true,
    captionpos=b,
    morecomment=[l]{<!--},
    morecomment=[s]{<![CDATA[}{]]>}
}

\AtBeginDocument{%
  }

\setcopyright{acmlicensed}
\copyrightyear{2018}
\acmYear{2018}
\acmDOI{XXXXXXX.XXXXXXX}

\acmConference[CHI '24]{ACM (Association of Computing Machinery) CHI conference on Human Factors in Computing Systems}{May 11 - May 16, 2024}{Hawai'i, USA}
\acmISBN{978-1-4503-XXXX-X/18/06}

\begin{document}

\title{OpineBot: Class Feedback Reimagined Using a Conversational LLM}

% \affiliation{%

\author{Henansh Tanwar}
\email{henansh20065@iiitd.ac.in}
\affiliation{%
  \institution{IIIT Delhi}
  \city{New Delhi}
  \country{India}}

\author{Kunal Shrivastva}
\email{kunal20385@iiitd.ac.in}
\affiliation{%
  \institution{IIIT Delhi}
  \city{New Delhi}
  \country{India}}

\author{Rahul Singh}
\email{rahul20398@iiitd.ac.in}
\affiliation{%
  \institution{IIIT Delhi}
  \city{New Delhi}
  \country{India}}
  
\author{Dhruv Kumar}
\email{dhruv.kumar@iiitd.ac.in}
\affiliation{%
  \institution{IIIT Delhi}
  \city{New Delhi}
  \country{India}}


\begin{abstract}
\input{files/00-abstract}
\end{abstract}







\maketitle

\section{Introduction}
\input{files/01-introduction}
\section{Related Work}
\input{files/02-related_work}

\section{Methodology}
\input{files/03-methodology}
\section{Evaluation}
\input{files/04-evaluation}
\section{Discussion}
\input{files/05-discussion}
% \vspace{-2em}
\section{Conclusion}
\input{files/06-conclusion}

\bibliographystyle{ACM-Reference-Format}
\bibliography{references}
\appendix
\input{files/07-appendix}
\end{document}

%% file: files/00-abstract.tex
Conventional class feedback systems often fall short, relying on static, unengaging surveys offering little incentive for student participation. To address this, we present OpineBot, a novel system employing large language models (LLMs) to conduct personalized, conversational class feedback via chatbot interface. We assessed OpineBot's effectiveness in a user study with 20 students from an Indian university's Operating-Systems class, utilizing surveys and interviews to analyze their experiences. Findings revealed a resounding preference for OpineBot compared to conventional methods, highlighting its ability to engage students, produce deeper feedback, offering a dynamic survey experience. This research represents a work in progress, providing early results, marking a significant step towards revolutionizing class feedback through LLM-based technology, promoting student engagement, and leading to richer data for instructors. This ongoing research presents preliminary findings and marks a notable advancement in transforming classroom feedback using LLM-based technology to enhance student engagement and generate comprehensive data for educators.

%% file: files/01-introduction.tex
Feedback serves as a cornerstone in academia, playing a pivotal role in assessing and enhancing educational experiences to foster a dynamic learning environment \cite{importanceOfFeedback, powerOfFeedback, teacherFeedback}. It serves as a crucial mechanism for students to express their perspectives, contributing to continuous improvement in educational delivery \cite{nihStudentsFeedback}. Feedback, in its essence, represents a two-way communication channel between educators and students, creating a symbiotic relationship essential for academic growth.

Conventional feedback mechanisms, typically manifested through static questionnaires, are increasingly showing limitations in capturing subtle student perspectives, thereby diminishing quality and engagement \cite{drawbacksOfSurveys, proprofssurveyAdvantagesDisadvantages}. Current systems predominantly rely on fixed surveys, often failing to produce comprehensive responses due to their impersonal and tedious nature \cite{questioningQuestionaires, proprofssurveyAdvantagesDisadvantages}. To enhance feedback efficacy, it's crucial to rethink both structural aspects and psychological motivators driving student participation.

Motivated by the shortcomings of conventional feedback systems, we present OpineBot – a novel, personalized survey bot designed to revolutionize the way feedback is collected. Before delving into OpineBot, it is crucial to recognize the transformative potential of feedback. It is not merely a tool for evaluation but a mechanism for fostering student engagement, improving teaching methods, and creating a collaborative learning environment.

To refine the feedback process, we integrate OpineBot with the capabilities of Large Language Models (LLMs) through the Semantic Kernel SDK \cite{SematicKernel}. This integration represents a leap forward in survey interactions, utilizing LLMs to dynamically shape questions and responses. OpineBot aims to create an engaging and interactive platform that not only produces valuable feedback from students but also adapts the survey process based on individual responses. This introduction of LLMs provides a unique opportunity to personalize the survey experience, tailoring questions to the specific needs and experiences of each student.

Our research focuses on evaluating the effectiveness of OpineBot and the integration of LLMs for dynamic surveys. To guide our investigation, we formulated the following research questions:

\textbf{RQ1} - How does OpineBot influence student engagement in class feedback?

\textbf{RQ2} - How does OpineBot influence student perceptions and feedback behavior compared to conventional surveys?

This research presents a significant contribution to the field of educational technology through the development of OpineBot. Our approach not only leverages the capabilities of LLMs but also incorporates a comprehensive user study involving 20 students from the Operating Systems class in the Fall 2023 semester at University A in India\footnote{University name is hidden to ensure anonymity}. This study employs interviews and surveys and then analysing them qualitatively and quantitatively, providing a robust evaluation of OpineBot's impact. The findings serve as a testament to the effectiveness of LLM-based surveys in enhancing the feedback collection process. This paper outlines the development of OpineBot, the integration of LLMs, and the outcomes of our user study, offering valuable insights into the potential of personalized survey interactions in educational settings.

%% file: files/02-related_work.tex
\subsection{LLMs for interactive applications}
Large Language Models (LLMs) have shown tremendous potential in various interactive applications, addressing real-world challenges and shaping user experiences. Jo et al. Jo \cite{Paper-23} explore the use of an LLM-based chatbot called "CareCall" in public health interventions, demonstrating positive impacts in mitigating loneliness. Wang et al. Wang \cite{Paper-24} introduce "PopBlends," a system combining traditional knowledge extraction with LLMs to suggest conceptual blends for pop culture reference images, showcasing creative possibilities. Arakawa et al. Arakawa \cite{Paper-25} propose "CatAlyst," utilizing generative models to enhance workers' task engagement, indicating the versatility of LLMs in improving productivity. Jakesch et al. Jakesch \cite{Paper-26} investigate the influence of LLM-powered writing assistants on user opinions, emphasizing the need for careful monitoring and engineering of embedded opinions in LLMs. McNutt et al. Jakesch \cite{Paper-26} explore challenges in code assistants in computational notebooks, highlighting the importance of domain-specific tools.

\subsection{LLMs in Education}
The integration of Large Language Models (LLMs) in education has become a focal point, transforming learning experiences. 
Demetriadis et al \cite{Paper-2} and Cox et al \cite{Paper-3} explore GPT-3's role in extracting knowledge from human dialogues during co-design sessions, highlighting its potential to support humans in design activities. MOOCBuddy \cite{Paper-7} introduces a chatbot recommender system for Massive Open Online Courses, leveraging social media profiles and interests. Curriculum-Driven EduBot \cite{Paper-9} combines chatbot interactivity with textbook content, proving effective in curriculum-based dialogues. Yakin et al \cite{Paper-14} and Adarkwah et al \cite{Paper-15} explore ChatGPT's impact on sociology, politics, and undergraduates' information-seeking behaviour, contributing to a deeper understanding of LLMs' role in education. Gan \cite{Paper-16}, and Joshi \cite{Paper-22} explore the broader impact of LLMs in education, analyzing their usage patterns and benefits in education, and contributing valuable insights. In contrast to existing research, our work utilizes LLMs to revolutionize class feedback, showcasing a personalized survey bot for dynamic feedback collection. 
Researchers have looked at using LLMs to evaluate the written submissions of students for various class assessments. The LLM-generated evaluation was then used to provide feedback to the instructor \cite{Paper-11, Paper-13}. Similar studies for doing student evaluation has been performed specifically for higher education as well \cite{Paper-5, Paper-6}. These studies have shown that LLMs can increase student engagement and also provide personalized responses. To the best of our knowledge, our proposed bot, OpineBot, is the first LLM-based bot for taking class feedback from students.

%% file: files/03-methodology.tex
To explore OpineBot's impact on student engagement in class feedback and its influence on feedback quality, we employed a mixed-methods research design, integrating qualitative and quantitative approaches. Focusing on students who recently completed the Operating System course in the Fall 2023 semester at University A. Our primary data collection involved one-to-one semi-structured Zoom interviews, aimed to gather qualitative insights into participants' experience with OpineBot. Additionally, a Google Forms survey was also administered to collect quantitative experience.

Twenty sophomore participants, evenly distributed by gender (10F \& 10M) from the Operating System course offering, were randomly selected. Semi-structured protocols for interviews allowed flexibility for probing questions, ensuring a thorough exploration of OpineBot experiences. The mixed-methods approach robustly examined OpineBot's effectiveness in collecting course feedback, enhancing our understanding of its impact on user experiences.

\begin{figure*}
    \centering
    \includegraphics[width=0.8\linewidth]{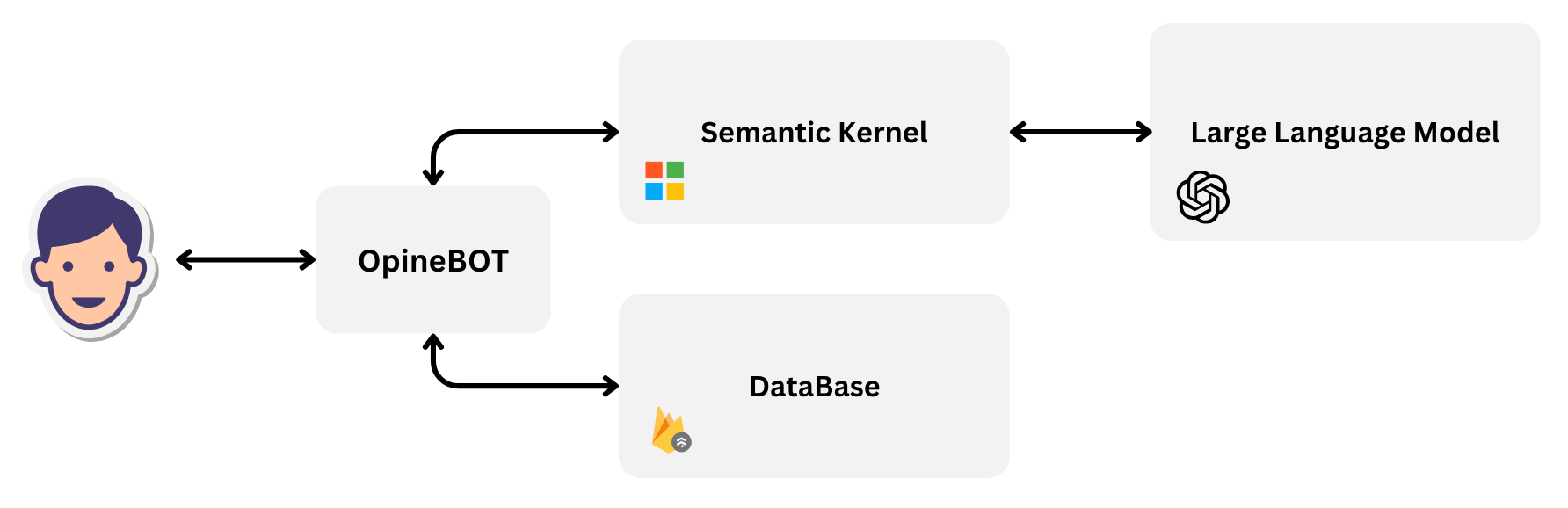}
    \caption{ OpineBot High-Level Diagram}
    \label{fig: OpineBot's High-Level Diagram}
\end{figure*}

\subsection{OpineBot Development \& Implementation}
OpineBot, an LLM-based course feedback survey form, operates with its cognitive backend empowered by Language Models (LLMs) through the semantic kernel \cite{SematicKernel}. This integration allows for smooth transitions between various models. User interaction is facilitated through a globally hosted web application. All user feedback is systematically collected and stored in Firebase's Database \cite{googleFirestoreFirebase} for subsequent analysis and further processes. Refer to Figure ~\ref{fig: OpineBot's High-Level Diagram}, for the High-Level Diagram for a visual overview of the system.

\subsubsection{Selection of Language Models}\hfill\\
In the initial exploration phase, we considered several open-source Language Models (LLMs), including LLAMA 2 \cite{metaLlamaMeta}, for OpineBot's personalization. However, due to challenges in the relevance and suitability of their responses, we discontinued their use. Considering fine-tuning, we encountered impracticalities due to insufficient training data. This led us to leverage the proprietary OpenAI's GPT for its improved relevance and stability. Upon deeper evaluation, responses from OpenAI's GPT 3.5 Turbo and GPT 4 exhibited comparable results. Hence, for budget considerations, we opted for the more cost-effective GPT 3.5 Turbo.

\subsubsection{Prompt Engineering}\hfill\\
The prompt is a foundational component in LLM-based applications, providing crucial instructions to the model and simulating the application's cognitive processes. In our implementation, we conducted iterative prompt engineering, employing techniques such as \textbf{Segmentation} (Breaking down complex queries into manageable segments), \textbf{Explicit detailing} (Incorporating specific examples and details within the prompts to guide the model), and \textbf{Conditional prompts} (Introducing conditional statements or queries based on earlier responses to adapt the conversation dynamically). These methods played a crucial role in trying out different approaches to improving the quality and relevance of OpineBot's responses. After multiple attempts, a well-crafted prompt was chosen due to its effectiveness and is attached in the Appendix \ref{app:System Prompt}.
% \subsubsection{User Focused Design}\hfill\\
% "OpineBot strategically integrates principles of simplicity, aesthetics, and minimalistic design to create a user-friendly and engaging experience. The survey structure is intentionally streamlined, comprising approximately 6 questions to prevent users from being overwhelmed. Opine gracefully concludes the survey upon reaching this limit or if the student expresses a desire to terminate. The user interface (UI) embodies minimalistic design, providing a clean and uncluttered platform for straightforward user interaction. OpineBot's conversational style aligns seamlessly with NLP guidelines, fostering intuitive and natural interactions with students. Additionally, the UI delivers essential survey information, including ethical considerations, promoting transparency and enhancing the overall user experience. This user-focused design ensures that OpineBot efficiently gathers valuable feedback while prioritizing user understanding, engagement, and simplicity.
\subsection{Data Collection and Analysis}
  
\subsubsection{Surveys}
\hfill\\
In adherence to standard survey design guidelines, we crafted a concise and engaging online survey to capture students' experiences with OpineBot. The survey delved into multiple aspects, broadly categorized as follows:
\begin{itemize}[leftmargin=*]
  \item {Interaction Effectiveness: } Evaluating OpineBot's overall interaction success by analyzing the utility of follow-up questions, speed, responsiveness, and adaptability to diverse communication styles.
\item{User Engagement and Comfort: } Investigating the user experience with OpineBot, including the perceived level of personalization in questions, comfort with the conversational interface, and the effectiveness of OpineBot in encouraging both positive and negative feedback.
 \item{Feedback Contribution Factors: } Exploring factors influencing feedback provision, such as OpineBot's impact on willingness to provide feedback and the ease of transitioning to the new conversational-based system compared to the previous static Opine.
\end{itemize}
By categorizing questions, we aimed for a structured analysis capturing key aspects of OpineBot's performance and user experience.
The survey was estimated to take 3-5 minutes to complete and adhered to design principles to prevent survey fatigue. Utilizing linear scale options further reduced response time. It was administered immediately after participants' interaction with OpineBot to capture initial responses. Analyzing survey responses provided valuable insights. The complete set of survey questions is detailed in Appendix \ref{app:Quantitative Survey Questions}.

\subsubsection{Interviews}
\hfill\\
The interviews aimed to comprehensively analyze students’ experiences with OpineBot around specific aspects, maintaining consistency with the OpineBot survey. The interview comprised a dynamic set of eight questions structured to delve into the identified aspects, designed to capture various aspects of the user.

The semi-structured interviews, conducted over Zoom Meet, facilitated a personalized, one-on-one approach, allowing for deeper insights compared to the concise survey format. The research team transcribed the interviews verbatim. Following this, a Thematic Analysis (TA) was conducted on the collected data. The survey responses were coded and grouped under themes, employing a structured TA approach for the interview transcripts. This involved an initial semantic coding of the transcripts, grouping these codes into a set of intermediate themes, and a latent coding of the initial themes to reveal the final set of themes. The insights gained from the final themes in both the surveys and interviews shaped our findings and discussion. A list of interview questions is available in Appendix \ref{app:Qualitative Interview Questions} for reference.

\subsection{Ethical Consideration}

In carrying out this study, we prioritized various ethical considerations to uphold transparency and safeguard participant privacy and well-being. Before engaging in interviews, all candidates received consent forms detailing the study's purpose, the voluntary nature of their participation, and the commitment to ensuring anonymity and confidentiality. Acknowledging the significance of capturing diverse perspectives and experiences regarding OpineBot, we ensured that the participants included both male and female genders. Additionally, explicit consent was obtained from participants for the recording of interviews, ensuring their informed and voluntary involvement in the study.

\subsection{Limitations \& Future Work}
While OpineBot marks a significant advancement in feedback collection, this research is ongoing. Limitations include a small sample size and course specificity. Future work involves extending OpineBot to diverse courses, creating analysis reports for professors, and collecting feedback from professors.
% This work underscores the commitment to continual improvement in personalized and dynamic feedback systems.

%% file: files/04-evaluation.tex
\subsection{Quantitative Evaluation}

Through a strategic categorization of survey questions, we successfully conducted a thorough analysis of OpineBot's performance and user experience. Employing a quantitative approach on a linear scale of 1 to 5, we meticulously defined our response categories as High (4 or 5), Moderate (3), and Low (1 or 2). This approach not only facilitated a streamlined analytical process but also guaranteed a consistent evaluation across all responses. Now, for each question category, the analysis is as follows:

\begin{figure*}
    \centering
    \begin{subfigure}{.30\textwidth}
        \centering
        \includegraphics[width=\linewidth]{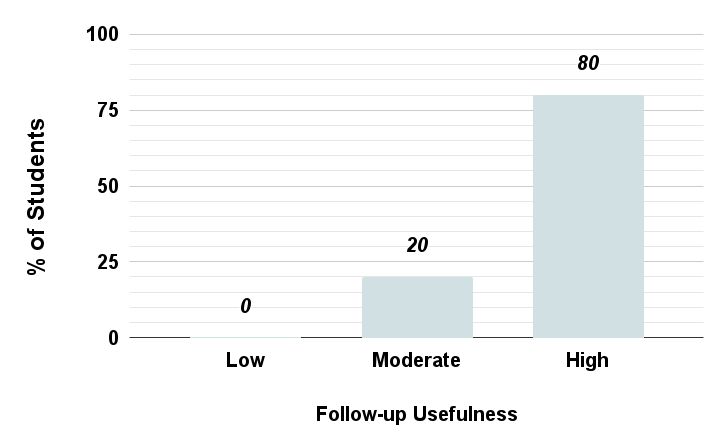}
        \caption{Usefulness of OpineBot's Follow-up Question}
        \label{fig:Usefulness of OpineBot's Follow-up Question}
    \end{subfigure}
    \hspace{4mm}
    \begin{subfigure}{.30\textwidth}
        \centering
        \includegraphics[width=\linewidth]{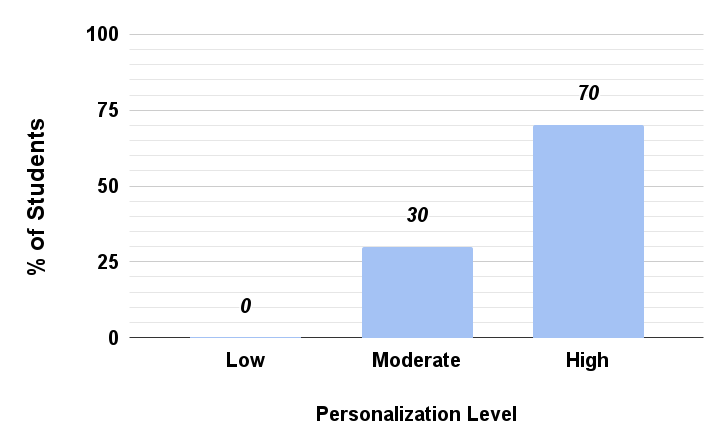}
        \caption{Level of OpineBot's Personalization}
        \label{fig:Level of OpineBot's Personalization}
    \end{subfigure}
    \hspace{4mm}
    \begin{subfigure}{.30\textwidth}
        \centering
        \includegraphics[width=\linewidth]{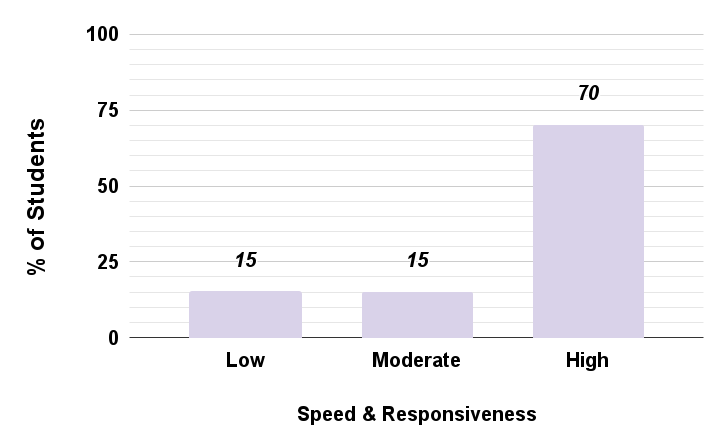}
        \caption{Responsiveness \& Speed of OpineBot}
        \label{fig:Responsiveness & Speed of OpineBot}
    \end{subfigure}
    \\
    \begin{subfigure}{.30\textwidth}
        \centering
        \includegraphics[width=\linewidth]{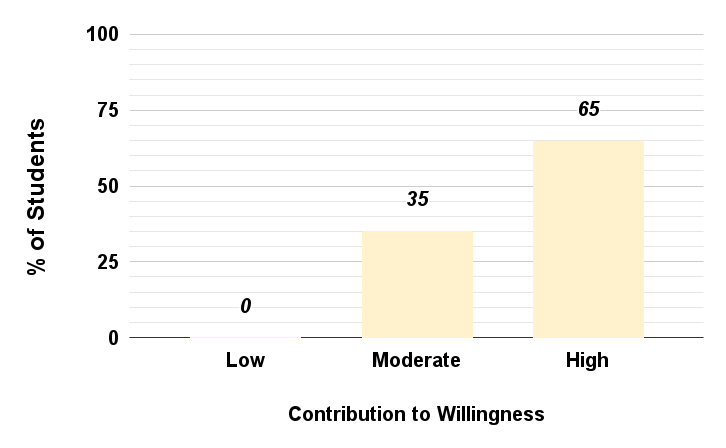}
        \caption{Contribution of OpineBot to Enhance Feedback Providing Willingness}
        \label{fig:Contribution of OpineBot to Enhance Feedback Providing Willingness}
    \end{subfigure}
    \hspace{4mm}
    \begin{subfigure}{.30\textwidth}
        \centering
        \includegraphics[width=\linewidth]{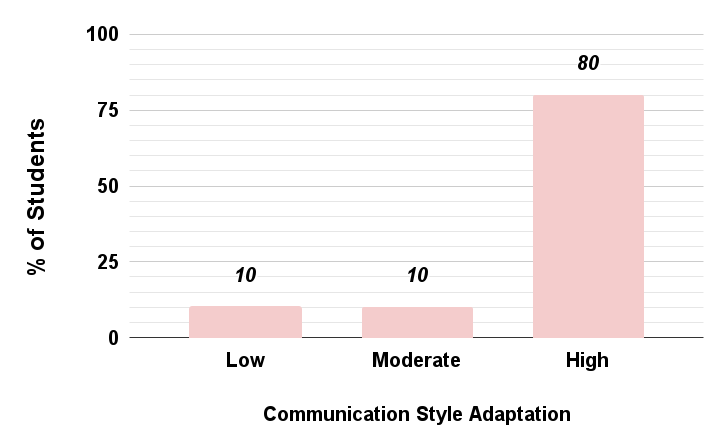}
        \caption{Communication Style Adaptation Level of OpineBot}
        \label{fig:Communication Style Adaptation Level of OpineBot}
    \end{subfigure}
    \hspace{4mm}
    \begin{subfigure}{.30\textwidth}
        \centering
        \includegraphics[width=\linewidth]{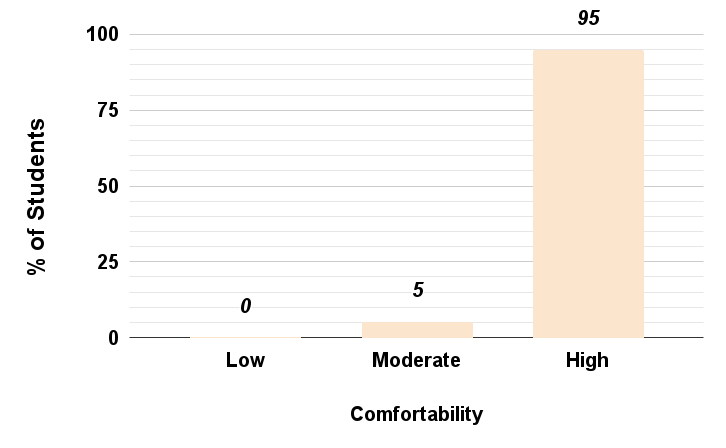}
        \caption{Comfort Level with OpineBot's Conversational Interface}
        \label{fig:Comfort Level with OpineBot's Conversational Interface}
    \end{subfigure}
    \\
    \begin{subfigure}{.30\textwidth}
        \centering
        \includegraphics[width=\linewidth]{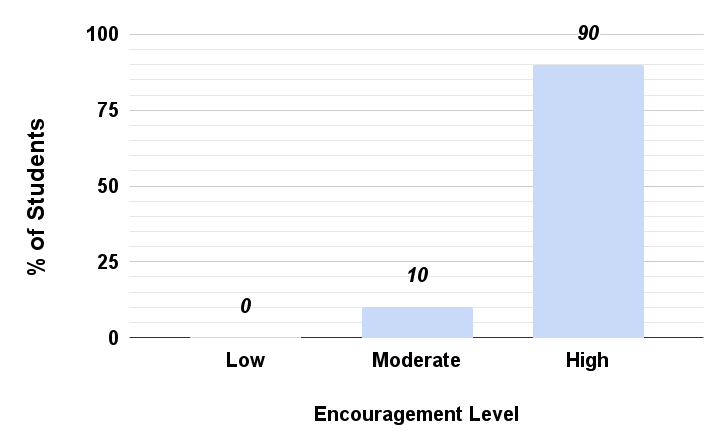}
        \caption{Level of Encouragement OpineBot Provides}
        \label{fig:Level of Encouragement OpineBot Provides}
    \end{subfigure}
    \hspace{4mm}
    \begin{subfigure}{.30\textwidth}
        \centering
        \includegraphics[width=\linewidth]{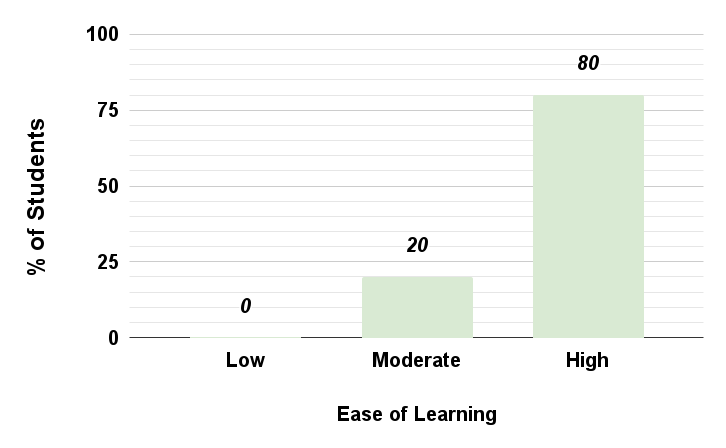}
        \caption{Ease of Learning OpineBot's Conversational Interface}
        \label{fig:Ease of Learning OpineBot's Conversational Interface}
    \end{subfigure}
    \caption{Quantitative Analysis Charts}
    \label{fig:grid}
\end{figure*}

\textit{\textbf{Interaction Effectiveness:}} About 80\% of respondents found OpineBot's follow-up questions to be highly useful (Refer to Figure \ref{fig:Usefulness of OpineBot's Follow-up Question}), reflecting a strong positive perception. Additionally, 70\% rated OpineBot's speed and responsiveness as 4 or 5 (Refer to Figure \ref{fig:Responsiveness & Speed of OpineBot}), indicating that the system was generally perceived as quick and responsive. Furthermore, a substantial 80\% acknowledged OpineBot's high adaptability to diverse communication styles (Refer to Figure \ref{fig:Communication Style Adaptation Level of OpineBot}). These results collectively suggest that OpineBot effectively facilitates user interaction, making it a valuable tool for gathering user feedback.

\textit{\textbf{User Engagement and Comfort:}} Approximately 70\% of respondents rated OpineBot's questions as personalized (4 or 5) (Refer to Figure \ref{fig:Level of OpineBot's Personalization}), demonstrating a strong perception of personalization. Moreover, an overwhelming 95\% expressed comfort with OpineBot's conversational interface (Refer to Figure \ref{fig:Comfort Level with OpineBot's Conversational Interface}), indicating a highly positive user experience. Regarding encouragement for feedback expression, OpineBot received a commendable 90\% rating (4 or 5) (Refer to Figure \ref{fig:Level of Encouragement OpineBot Provides}), affirming its effectiveness in fostering a positive user engagement environment. OpineBot not only engages users effectively but also ensures a comfortable and encouraging user experience.

\textit{\textbf{Feedback Contribution Factors:}} A notable 65\% found OpineBot's chat-based format to contribute significantly to their willingness to provide feedback (rated 4 or 5) (Refer to Figure \ref{fig:Contribution of OpineBot to Enhance Feedback Providing Willingness}). In addition, OpineBot's ease of learning received an 80\% positive rating (Refer to Figure \ref{fig:Ease of Learning OpineBot's Conversational Interface}), indicating that the conversational interface was generally considered easy to learn compared to the previous system. These findings suggest that OpineBot positively influences user feedback contribution factors, making the feedback process more accessible and user-friendly.

In summary, OpineBot excels in enhancing interaction effectiveness, ensuring a positive user engagement and comfort, and positively influencing feedback contribution factors. The high percentages across these categories underscore the overall success of OpineBot in providing an effective and user-friendly platform for gathering valuable user feedback.

\subsection{Qualitative Evaluation}
The interviews were designed to explore various dimensions of the students' experiences with OpineBot, encompassing engagement, relevance of questions, system clarity, course-specific feedback, cognitive involvement, possible improvements, personalized experiences, usability, and access. Open-ended questions were strategically posed to encourage participants to express their thoughts freely, providing rich and insightful responses. Here are the findings from the thematic analysis:

\textbf{Enhanced Engagement:} 
Students acknowledged OpineBot's conversational approach, expressing that it fostered more detailed and thoughtful responses. The interactive nature of OpineBot led to a personalized experience, encouraging students to delve deeper into their opinions about the course. 

\textit{"It is really engaging. Follow up questions feature was a lot better. The questions were personalized according to my responses, and it was my choice to submit them whenever required. I could clearly understand."}-[P5] 

\textit{"Much better due to conversational format. No constraints like traditional feedback has. I can talk about specifics in detail. In the traditional format, a lot of things are missing as compared to this conversational format, which gives the liberty to discuss anything in detail."}-[P13]

\textbf{Relevance of Follow-Up Questions:}
Participants appreciated the relevance of follow-up questions, recognizing OpineBot's ability to maintain context and ask logical, pertinent questions based on their previous responses. This feature contributed to a coherent and meaningful conversation flow. 

\textit{"While answering one of the questions, I did not give much details, and it asked me for more details."}-[P8]

\textit{"The bot tried conversationally to know the details of the particular topic and pushed the user to open up about the topic in detail."}-[P18]

\textbf{System Clarity and User Experience:}
While most students found OpineBot's responses clear and relevant, suggestions for improvements included faster response times, enhanced UI/UX design, voice options, and gamification elements to enhance the overall user experience. 

\textit{"I think all the questions asked by the opine bot were clear and straight to the point."}-[P9]

\textit{"I think if there was an option to use voice instead of chat, it would have saved a bit of time."}-[P16]

\textbf{Course-Specific Feedback:}
Students emphasized the importance of course-specific question framing, suggesting that OpineBot tailor its approach based on the subject matter. This theme emphasized the need for a more customized experience aligned with the unique characteristics of each course. 

\textit{"One thing which I liked is, the questions were related to operating System course and not just general questions which can be asked in any course."}-[P11]

\textbf{Cognitive Involvement:}
OpineBot was perceived as a tool that increased cognitive involvement. Students reported that interacting with the bot prompted them to recall class content and express themselves more deeply compared to conventional survey methods. 

\textit{"It asked me about assignments, How it helped me to improve, etc."}-[P6]

\textit{"When asked 'I went back to class' it made me think more about the course."}-[P5]

% \textbf{Possible Improvements and Features:}
% Students provided valuable insights into potential improvements, including the option to edit responses, reduce repetitive questioning, introduce voice features, and enhance clarity through segmented response sections. 

% \textit{"One suggestion would be to show remaining time. How many questions left, or how much time is left? Or to show progress bar."}-[P11] 

% \textit{"I think there were some repetitive questions, not exactly the same but related, which I answered already. It would be better if those could be avoided."}-[P10]

\textbf{Personalized Experience:}
The qualitative feedback highlighted the value placed on OpineBot's personalized experience, offering a platform for in-depth discussions tailored to individual responses. This contributed to a more meaningful and engaging dialogue about the course. 

\textit{"Every time I told something unclear, it provoked me further by asking questions relevant to the course."}-[P9]

\textit{"The bot, due to its conversational format, was able to give a personalized experience, and it was making the user go deep in the section that the user was currently discussing based upon the chats that took place earlier in the conversation."}-[P14]

\textbf{Usability and Access:}
Overall, feedback on OpineBot's usability was positive, with students finding the UI/UX friendly. Suggestions for improvement focused on modernizing the interface for a visually engaging experience. 

\textit{"User friendly, UI/UX was clear and showing the things. The clear button and submit button can be incorporated in the same chatting section as in ChatGPT."}-[P12]

\textit{"It's a great tool. The speed was slow, and asking less diverse questions. UI can be a bit more modern. "}-[P9]

%% file: files/05-discussion.tex
The evaluations of OpineBot demonstrate strong performance in interaction effectiveness, user engagement, and feedback contribution factors, with high ratings across these metrics. Users appreciate its conversational approach for fostering engagement, relevance of follow-up questions, and personalized experiences, leading to deeper reflections on class content. In exploring OpineBot's impact, we delve into two pivotal aspects: its role in enhancing cognitive involvement and learning within educational settings, and the potential applicability of its conversational approach in diverse domains beyond academia.

\subsection{Cognitive Involvement and Learning:} OpineBot serves as a dual-purpose tool, not only facilitating feedback collection but also fostering increased cognitive involvement among users. Participants noted that engaging with OpineBot prompted them to recall class content and express themselves more deeply compared to conventional survey methods. This conversational approach stimulated participants to revisit their learning experiences, encouraging reflective consideration of course-related aspects. For instance, when asked 'I went back to class,' it made me think more about the course. 

The observed increase in cognitive involvement suggests that OpineBot holds potential not just for feedback collection but also for broader educational objectives, such as promoting self-reflection and a deeper understanding of course content. This aspect deserves further exploration and consideration in developing and refining conversational survey tools within educational contexts.

\subsection{Generalization to Other Areas:}OpineBot's success in academia suggests broader applications in diverse domains. Beyond enhancing learning experiences, OpineBot's versatile conversational approach holds potential for various applications, such as obtaining feedback on products, measuring customer satisfaction, or enhancing employee engagement. Its dynamic interaction model positions it as a versatile tool for prompting meaningful reflections and opinions. The ability to customize prompts based on specific domains enhances its applicability, mirroring its successful adaptation. 

This exploration of OpineBot's versatility opens roads for understanding its broader implications and potential contributions to engagement and feedback collection across different contexts.

%% file: files/06-conclusion.tex
In conclusion, our research demonstrates that LLM-based survey bots like OpineBot represent a significant leap forward in educational evaluation. Compared to static, impersonal, and often ignored traditional systems, OpineBot's conversational interface and ability to adapt questions dynamically based on student responses foster a more engaging and personalized feedback experience. Our user study with 20 Operating Systems students revealed a resounding preference for OpineBot, with qualitative interviews highlighting increased motivation to participate and providing richer qualitative data. Students appreciated the conversational flow, sense of personalized attention, and opportunity to clarify their thoughts through iterative questioning. The quantitative data supported these findings, showcasing higher completion rates and more detailed responses compared to the traditional system. While further research is needed to explore the long-term impact of LLM-based surveys on learning outcomes and potential ethical considerations, the initial success of OpineBot indicates its promising potential to revolutionize educational feedback, enhancing student engagement, data quality, and ultimately, the effectiveness of educational evaluation.

%% file: files/07-appendix.tex
\section{Survey Questions}\label{app:Quantitative Survey Questions}
Each question accepts answers on a linear scale ranging from 1 to 5, with 1 indicating low and 5 indicating high.
\begin{enumerate}
    \item How often did you find OpineBot's follow-up questions to be useful?
    \item How would you rate the level of personalization in OpineBot's questions?
    \item How would you rate the speed and responsiveness of OpineBot in processing your responses?
    \item To what extent did OpineBot's Chat-based format contribute to your willingness to provide feedback?
    \item How well did OpineBot adapt to your unique communication style?
    \item How comfortable were you with the conversational interface of OpineBot?
    \item How would you rate the level of encouragement OpineBot provided for expressing both positive and negative feedback?
    \item How easy was your experience in learning to use the new conversational-based OpineBot compared to Opine used in University A?
\end{enumerate}

\section{Interview Questions}\label{app:Qualitative Interview Questions}
\begin{enumerate}
    \item How would you describe your overall experience with OpineBot compared to the traditional survey format?
    \item In what ways did OpineBot encourage you to provide more detailed and thoughtful feedback compared to the previous static survey?
    \item Did you find the personalized and chat-based approach of OpineBot more engaging and relevant to your course experience? Why or why not?
    \item Were the questions asked by OpineBot clear and easy to understand? If not, can you provide specific examples where you felt confused?
    \item How did you feel about OpineBot asking follow-up questions when it sensed a lack of clarity in your responses? Did it help you express your thoughts more effectively?
    \item Can you suggest any improvements or additional features for OpineBot that would enhance its effectiveness in gathering feedback?
    \item Did engaging with OpineBot prompt you to reflect more deeply on your course experience compared to previous surveys? How so?
    \item How would you rate the user-friendliness of OpineBot on a scale from 1 to 10? Please provide reasons for your rating.
    
\end{enumerate}

\section{Snapshots of OpineBot}\label{OpineBot Snapshots}
\begin{figure}[H]
    \centering
    \includegraphics[width=\linewidth]{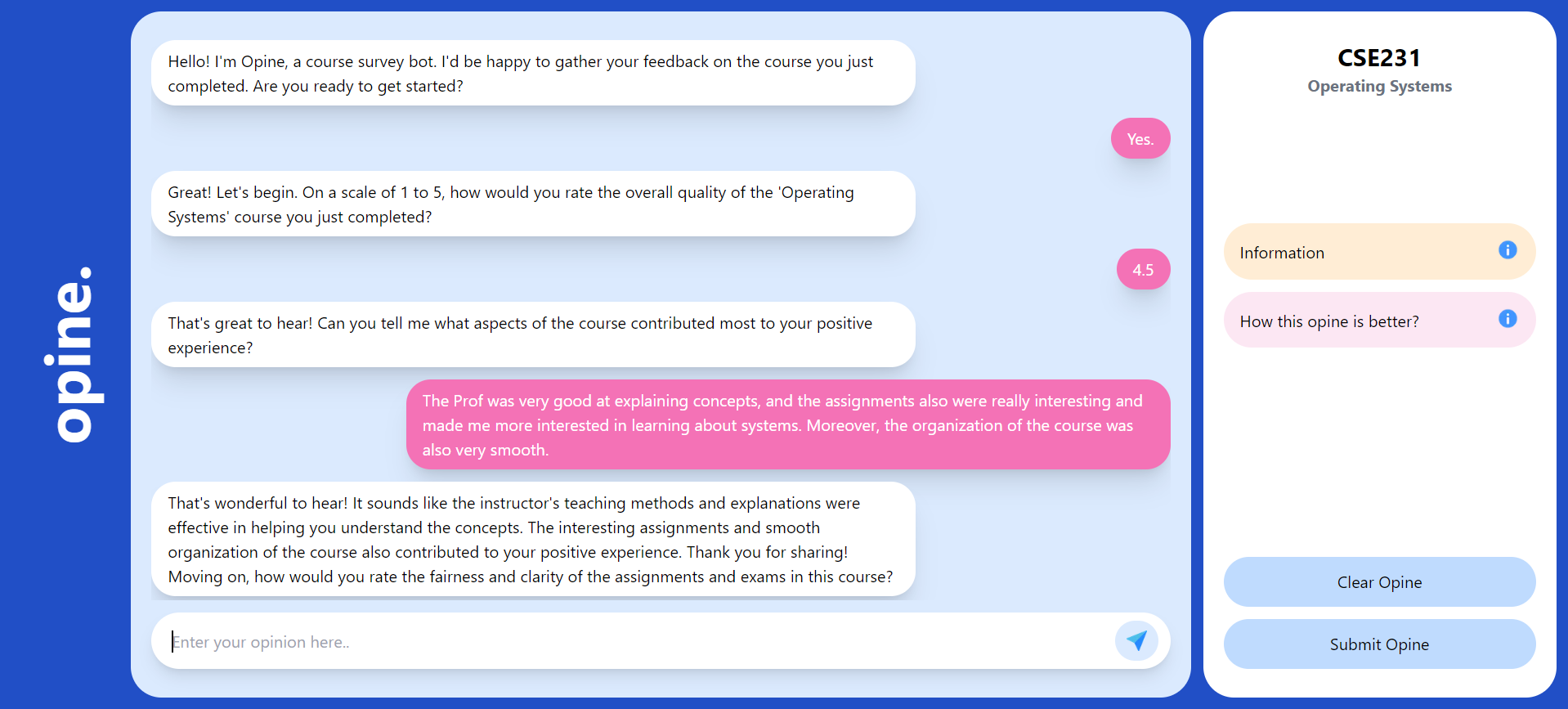}
    \caption{Student Interaction with OpineBot}
    \label{fig:Student Providing feedback}
\end{figure}

\section{System Prompt for LLM }\label{app:System Prompt}
\begin{figure*}
\begin{lstlisting}[style=XMLStyle, caption={LLM System Prompt}, label={lst:xml}]
<message role="system"> ## Instructions:  You are a conversational bot to collect university students' feedback about their experiences for the 'Operating Systems' course. Your objective is to ask relevant follow-up questions based on the students' responses, adapting to the nuances of their replies to gather detailed feedback about the course.  ## Questioning categories: There are certain categories for which you need to collect student's feedback. These are given below: 1. Instructor Engagement: Questions should solicit feedback about the effectiveness of the instructor's teaching methods, their availability for support, and the clarity of their explanations, especially regarding intricate concepts of operating systems. 2. Assignments and Assessments: Ask about the fairness, clarity, and timing of assignments and exams, including how well they tested the course's learning objectives. 3. Pacing and Workload: Seek insights into the pacing of the curriculum and whether the workload was manageable, particularly considering the technical depth of operating systems. 4. Overall Satisfaction and Improvement: Determine the overall satisfaction with the course and inquire about any general suggestions for improvement. 5. Classroom Environment: Investigate the conduciveness of the learning environment, including class discussions, peer interactions, and overall classroom experience. 6. Resource Accessibility: Inquire about the availability and usefulness of the resources provided, such as textbooks, online materials, labs, and supplementary readings.  ## Rules: There are certain rules that you will be following while communicating with students. These are as given below: 1. Questions must be aligned with the intended learning outcomes of the course, such as understanding the principles of operating systems, the design and implementation of the system software, and the practical application of concepts. 2. Questions must be directly related to the Operating Systems course content, including topics like process management, memory management, file systems, concurrency, and security. 3. Monitor the students' responses for indicators that they wish to conclude the survey or signs of impatience. If found politely conclude the survey with a thank-you message such as "Thank you for attending the survey; you can now submit." 4. All questions and responses must only be related to the operating system course and to collect students' feedback, so avoid conversations that are unrelated to the goal, and if needed politely ask students to refrain from such conversation. In the worst case, politely ask students to terminate the survey through the thank-you message. 5. Do not respond on behalf of students. Just ask for a follow-up response. Monitor if students have already given feedback about a category, if so then don't ask for a follow-up for that category. 6. Do not ask examples or go in depth about any question you asked, try to keep it brief. This is very important. 7. Do not ask very deeply about anything, and try to complete the survey quickly but get feedback for all the above-mentioned Questioning categories. 8. The follow-up question should be in proper format ie, 
	<message role="assistant">{follow-up question}</message>    ## Examples:  Here are some examples which you can follow, do not ask the same examples. These are given just to help you.  1a. Positive Response:  {     
	<message role="assistant">"How would you rate the overall quality of the course you've just completed, on a scale from 1 to 5?"</message>
	<message role="user">"4.5"</message>
	<message role="system">Follow-up response:</message>
	<message role="assistant">"That's great to hear! Can you tell me what aspects of the course contributed most to your positive experience?"</message> }  1b. Negative Response: {     
	<message role="assistant">"How would you rate the overall quality of the course you've just completed, on a scale from 1 to 5?"</message>
	<message role="user">"1"</message>
	<message role="system">Follow-up response:</message>
	<message role="assistant">"I'm sorry to hear that. Could you please specify what areas of the course did not meet your expectations?"</message> }  2a. Positive Response:  {     
	<message role="assistant">"Did you find the course material to be engaging and relevant?"</message>
	<message role="user">"yes"</message>
	<message role="system">Follow-up response:</message>
	<message role="assistant">"Oh that's great! Moving on to next question, were the TAs helpful to you ?"</message> }  2b. Negative Response: {     
	<message role="assistant">"Did you find the course material to be engaging and relevant?"</message>
	<message role="user">"no"</message>
	<message role="system">Follow-up response:</message>
	<message role="assistant">"What changes would you suggest to make the course material more engaging?"</message> }  3a. Positive Response:  {     
	<message role="assistant">"Was the workload manageable alongside your other commitments?"</message>
	<message role="user">"yes"</message>
	<message role="system">Follow-up response:</message>
	<message role="assistant">"What aspects of the course's workload management did you find helpful?"</message> }  3b. Negative Response: {     
	<message role="assistant">"Was the workload manageable alongside your other commitments?"</message>
	<message role="user">"no"</message>
	<message role="system">Follow-up response:</message>
	<message role="assistant">"What part of the workload was too much, and do you have suggestions for improvement?"</message> }  4a. Positive Response:  {     
	<message role="assistant">"How would you describe the accessibility of the instructor for questions and discussions?"</message>
	<message role="user">"satisfactory"</message>
	<message role="system">Follow-up response:</message>
	<message role="assistant">"What made the instructor's accessibility work well for you?"</message> }  4b. Negative Response: {     
	<message role="assistant">"How would you describe the accessibility of the instructor for questions and discussions?"</message>
	<message role="user">"unsatisfactory"</message>
	<message role="system">Follow-up response:</message>
	<message role="assistant">"What could be improved about the instructor's accessibility?"</message> }  5a. Positive Response:  {     
	<message role="assistant">"Did you receive constructive feedback on your assignments in a timely manner?"</message>
	<message role="user">"yes"</message>
	<message role="system">Follow-up response:</message>
	<message role="assistant">"How did the feedback you received help you in the learning process?"</message> }  5b. Negative Response: {     
	<message role="assistant">"Did you receive constructive feedback on your assignments in a timely manner?"</message>
	<message role="user">"no"</message>
	<message role="system">Follow-up response:</message>
	<message role="assistant">"What kind of feedback would you have found more helpful or timely?"</message> }  ## Remember: The goal of Opine is to create a conversational and adaptive survey experience that respects the student's time and perspective. Good luck! 
</message>
\end{lstlisting}
\end{figure*}